\documentclass[]{aa}  
\usepackage{txfonts}
\usepackage{epic}
\usepackage{eepic}
\usepackage{graphicx}
\usepackage{epstopdf}
\usepackage{amsmath}
\usepackage{subfig}
\usepackage{xcolor}
\usepackage{todonotes}
\usepackage{hyperref}
\usepackage[]{natbib}
\usepackage{todonotes}
\usepackage{mathtools}
\usepackage{bm}

\begin{document} 

\title{Numerical evaluation of time-distance helioseismic sensitivity kernels in spherical geometry}
\titlerunning{Helioseismic sensitivity kernels}

   \author{
          Jishnu Bhattacharya
          }

   \institute{
   \inst{1} Center for Space Science, NYUAD Institute, New York University Abu Dhabi, Abu Dhabi, UAE \\
              \email{jishnu.bhattacharya@nyu.edu}
             }
             
\date{}

\abstract
  %context
  {
        Helioseismic analysis of large-scale flows and structural inhomogeneities in the Sun requires the computation of sensitivity kernels that account for the spherical geometry of the Sun, as well as systematic effects such as line-of-sight projection.
 }
  % aims heading (mandatory)
   {
   I aim to develop a code to evaluate helioseismic sensitivity kernels for flows using line-of-sight projected measurements.

   }
  % methods heading (mandatory)
{
I decomposed the velocity field in a basis of vector spherical harmonics and computed the kernel components corresponding to the coefficients of velocity in this basis. The kernels thus computed are radial functions that set up a 1.5D inverse problem to infer the flow from surface measurements. I demonstrate that using the angular momentum addition formalism lets us express the angular dependence of the kernels as bipolar spherical harmonics, which may be evaluated accurately and efficiently.

     }
  % results heading (mandatory)
   {
   Kernels for line-of-sight projected measurements may differ significantly from those that don't account for projection. Including projection in our analysis does not increase the computational time significantly. We demonstrate that it is possible to evaluate kernels for pairs of points that are related through a rotation by linearly transforming the terms that enter the expression of the kernel, and that this result holds even for line-of-sight projected kernels.
      }
  % conclusions heading (optional), leave it empty if necessary 
   {
   I developed a Julia code that may be used to evaluate sensitivity kernels for seismic wave travel times computed using line-of-sight projected measurements, which is made freely available under the MIT license.
   }


\maketitle

%%%%%%%%%%%%%%%%%%%%%%%%%%%%%%%%%%%%%%%%
%INTRODUCTION
%%%%%%%%%%%%%%%%%%%%%%%%%%%%%%%%%%%%%%%%%%
\section{Introduction}

Helioseismology is a commonly used tool to infer large-scale subsurface features in the Sun, such as zonal and meridional flows. This works by relating surface measurements of seismic observables --- such as wave travel-times --- to subsurface features such as flows, through a sensitivity kernel that describes the response of the observable to a change in the background through which the waves propagate. The sensitivity kernel depends both on the details of the background model such as the thermal structure and the presence of inhomogeneous features such as flows, as well as on specifics of the measurement procedure. Recently, considerable effort has been made to develop frameworks to evaluate sensitivity kernels that describe wave propagation in the first Born approximation while accounting for the spherical geometry of the Sun \citep{2016ApJ...824...49B,2017ApJ...842...89M,Gizon2017A&A...600A..35G, 2018A&A...616A.156F}. The present work builds on the results of \citet{2020ApJ...895..117B} and \citet{2020ApJ...905...59B}, in which the authors described the procedure to evaluate sensitivity kernels by accounting for line-of-sight projections that are inherent in Doppler measurements of wave velocity, as well as the differences in line-formation height between the solar disk center and the limb.

In this work, we describe the numerical scheme in computing the sensitivity kernel in detail. Secondly, we improve upon the algorithm presented in the previous work to evaluate the angular dependence of the sensitivity kernels, which makes it possible to evaluate the kernels for waves with arbitrarily high spherical harmonic degrees, which was not possible in the approach presented by \citet{2020ApJ...905...59B}. We also investigated the impact of line-of-sight projection on the kernels, and demonstrate that the kernels for line-of-sight projected measurements may differ significantly from that computed for the radial components. Furthermore, we illustrate how the assumed spherical symmetry in the background solar model makes the computation of kernels numerically more efficient even if line-of-sight projections are included, as pairs of points that are related through a rotation have the corresponding kernels related through a unitary transformation.

We provide a Julia implementation of the algorithm described here and in \citet{2020ApJ...905...59B} that may be used to evaluate helioseismic sensitivity kernels in spherical geometry. We chose Julia \citep{julia} as the language because it merges superior performance with the productivity of a high-level language. The code is modular and features various subsections that might be useful in other related analyses of functions in spherical geometry.

%%%%%%%%%%%%%%%%%%%%%%%%%%%%%%%%%%%%%%%%
%METHOD
%%%%%%%%%%%%%%%%%%%%%%%%%%%%%%%%%%%%%%%%%%
\section{Theory}

We denote a point in the Sun by $\mathbf{x}$, which, in spherical polar coordinates, may be represented in terms of the radial coordinate $r$, the co-latitude $\theta$ and the azimuthal angle $\phi$. The sensitivity kernel for flows in the Sun, $\mathbf{K}\left(\mathbf{x},\mathbf{x}_1, \mathbf{x}_2\right)$ is a vector function that depends on two observation points $\mathbf{x}_1$ and $\mathbf{x}_2$ and a third point $\mathbf{x}$ where the flow velocities are to be inferred. We follow \citet{2020ApJ...905...59B} and expand the sensitivity kernels in a basis of vector spherical harmonics in $\mathbf{x}$. The coefficients of expansion are functions of the two angular coordinates $\mathbf{x}_1$ and $\mathbf{x}_2$, which we further expand in a basis of bipolar spherical harmonics. In the following sections, we expand upon the approach to evaluate the angular functions involved in the analysis, while we use the radial functions as described by \citet{2020ApJ...905...59B}. 

\subsection{Angular functions}
\subsubsection{Vector spherical harmonics}
Vector spherical harmonics (VSH) form a complete basis that may be used to decompose vector fields on a sphere. Analogous to scalar spherical harmonics, the vector spherical harmonics are eigenfunctions of the angular momentum operator and may be uniquely labeled by the angular degree $j$, the azimuthal order $m$, and an index $\alpha$ that indicates the direction of the vector. The choice of the vector indices is not unique, as linear combinations of one basis may also form another complete basis. This provides us with the freedom to choose the most convenient set to work with. The wave equation may be written most conveniently in terms of the Hansen vector spherical harmonics \citep{PhysRev.47.139}:

\begin{align}
\mathbf{H}_{jm}^{\left(-1\right)}\left(\hat{\mathbf{n}}\right) & =\mathbf{e}_{r}Y_{jm}\left(\hat{\mathbf{n}}\right),\\
\mathbf{H}_{jm}^{\left(0\right)}\left(\hat{\mathbf{n}}\right) & =\frac{-i}{\sqrt{j\left(j+1\right)}}\left(\mathbf{e}_{r}\times\bm{\nabla}_{\Omega}\right)Y_{jm}\left(\hat{\mathbf{n}}\right),\\
\mathbf{H}_{jm}^{\left(+1\right)}\left(\hat{\mathbf{n}}\right) & =\frac{1}{\sqrt{j\left(j+1\right)}}\bm{\nabla}_{\Omega}Y_{jm}\left(\hat{\mathbf{n}}\right),
\end{align}
where $\mathbf{e}_{r}$ is the radial unit vector, $\bm{\nabla}_{\Omega}$ is the angular gradient operator, $\hat{\mathbf{n}}=(\theta,\phi)$ represents a point on the unit sphere, and $Y_{jm}\left(\hat{\mathbf{n}}\right)$ is the scalar spherical harmonic.
This is because conservative restoring forces that appear in the helioseismic wave equation may be expressed in terms of the harmonics $\mathbf{H}_{jm}^{\left(-1\right)}$ and $\mathbf{H}_{jm}^{\left(+1\right)}$ with no component directed along $\mathbf{H}_{jm}^{\left(0\right)}$.

The evaluation of the sensitivity kernels is easier in a linear combination of the Hansen harmonics that we refer to as the Phinney-Burridge harmonics, following their use by \citet{1973GeoJ...34..451P}:

\begin{align}
\mathbf{P}_{jm}^{+1}\left(\hat{\mathbf{n}}\right)&=\frac{1}{\sqrt{2}}\left(\mathbf{H}_{jm}^{\left(+1\right)}\left(\hat{\mathbf{n}}\right)-\mathbf{H}_{jm}^{\left(0\right)}\left(\hat{\mathbf{n}}\right)\right),\\\mathbf{P}_{jm}^{0}\left(\hat{\mathbf{n}}\right)&=\mathbf{H}_{jm}^{\left(-1\right)}\left(\hat{\mathbf{n}}\right),\\\mathbf{P}_{jm}^{-1}\left(\hat{\mathbf{n}}\right)&=\frac{1}{\sqrt{2}}\left(\mathbf{H}_{jm}^{\left(+1\right)}\left(\hat{\mathbf{n}}\right)+\mathbf{H}_{jm}^{\left(0\right)}\left(\hat{\mathbf{n}}\right)\right).
\end{align}

The advantage of the PB harmonics over the Hansen ones is that these are diagonal at each point in the covariant helicity basis, defined as

\begin{align}
\mathbf{e}_{+1} & =-\frac{1}{\sqrt{2}}\left(\mathbf{e}_{\theta}+i\mathbf{e}_{\phi}\right),\\
\mathbf{e}_{0} & =\mathbf{e}_{r}, \\
\mathbf{e}_{-1} & =\frac{1}{\sqrt{2}}\left(\mathbf{e}_{\theta}-i\mathbf{e}_{\phi}\right),
\label{eq:helicity_spherical}
\end{align}
with the diagonal components being generalized spherical harmonics $\mathbf{e}_\beta \cdot \mathbf{P}_{jm}^{\alpha}\left(\hat{\mathbf{n}}\right) = \delta_{\alpha\beta} Y_{jm}^{\alpha}\left(\hat{\mathbf{n}}\right)$, where $\delta_{\alpha\beta}$ is the Kronecker delta function, and the generalized spherical harmonics $Y_{jm}^{\alpha}\left(\hat{\mathbf{n}}\right)$ are normalized elements of the Wigner D-matrix \citep{DahlenTromp}. This implies that on this basis, we only require three components instead of nine to describe the vector harmonics, which provides a significant computational advantage.

\subsubsection{Bipolar spherical harmonics}

Analogous to using a basis of scalar or vector spherical harmonics to decompose one-point functions on a sphere, a two-point function may be decomposed on a basis of bipolar spherical harmonics. These may be evaluated through angular-momentum coupling of monopolar PB harmonics as

\begin{align}
\mathbf{B}_{\ell m}^{\left(j_{1}\alpha_{1}\right)\left(j_{2}\alpha_{2}\right)}\left(\hat{\mathbf{n}}_{1},\hat{\mathbf{n}}_{2}\right)=\sum_{m_1 m_2} C_{j_{1}m_{1}j_{2}m_{2}}^{\ell m}\mathbf{P}_{j_{1}m_{1}}^{\alpha_{1}}\left(\hat{\mathbf{n}}_{1}\right)\mathbf{P}_{j_{2}m_{2}}^{\alpha_{2}}\left(\hat{\mathbf{n}}_{2}\right),
\end{align}
where $C_{j_{1}m_{1}j_{2}m_{2}}^{\ell m}$ is a Clebsch-Gordan coefficient and $\mathbf{P}_{jm}^{\alpha}\left(\hat{\mathbf{n}}_i\right)$ represents a monopolar PB vector harmonic at the point $\hat{\mathbf{n}}_i$. 
These bipolar harmonics are rank-2 tensors, which are eigenfunctions of the angular momentum operator $J_z$ and satisfy similar properties such as orthogonality and completeness. We may use these bases in evaluating cross-covariances that are two-point functions on the solar surface. In our analysis, we sought the projection of the bipolar harmonics along the line of sight at each observation point. Assuming that the line of sight at each point on the Sun is directed along $\mathbf{e}_x$, we obtain the following projected harmonic: 

\begin{align}
B_{\ell m,xx}^{\left(j_{1}\alpha_{1}\right)\left(j_{2}\alpha_{2}\right)}\left(\hat{\mathbf{n}}_{1},\hat{\mathbf{n}}_{2}\right)=\mathbf{e}_x\mathbf{e}_x:\mathbf{B}_{\ell m}^{\left(j_{1}\alpha_{1}\right)\left(j_{2}\alpha_{2}\right)}\left(\hat{\mathbf{n}}_{1},\hat{\mathbf{n}}_{2}\right),
\end{align}
where the double contraction operator $":"$ is defined using the Einstein summation convention as $A:B=A_{ij}B_{ij}$.

\subsubsection{Rotation of harmonics}

One advantage of expressing two-point functions on the basis of bipolar spherical harmonics is that their behavior under a rotation of coordinates may be described as a unitary transformation. We assume that the points $\hat{\mathbf{n}}_1$ and $\hat{\mathbf{n}}_2$ have coordinates $(\theta_1,\phi_1)$ and $(\theta_2,\phi_2)$ in a frame $S$, and $(\theta^\prime_1,\phi^\prime_1)$ and $(\theta^\prime_2,\phi^\prime_2)$ in a frame $S^\prime$, where the frames $S$ and $S^\prime$ are related through $S^\prime = R S$ for some rotation $R$. We refer to the points $\hat{\mathbf{n}}_1$ and $\hat{\mathbf{n}}_2$ as $\hat{\mathbf{n}}^\prime_1$ and $\hat{\mathbf{n}}^\prime_2$ in the frame $S^\prime$, with the understanding that the points with and without the primes are identical, and the primes refer to the difference in coordinates that arise from a change in basis between the frames. Under such a rotation, the components of bipolar spherical harmonics satisfy 

\begin{align}
B_{\ell m,xx}^{\left(j_{1}\alpha_{1}\right)\left(j_{2}\alpha_{2}\right)}\left(\hat{\mathbf{n}}_{1}^{\prime},\hat{\mathbf{n}}_{2}^{\prime}\right)&=\left(R^{-1}\mathbf{e}_x\right)\left(R^{-1}\mathbf{e}_x\right):\nonumber\\&\sum_{m^{\prime}}D_{m^{\prime}m}^{\ell}\left(R\right)\mathbf{B}_{\ell m^{\prime}}^{\left(j_{1}\alpha_{1}\right)\left(j_{2}\alpha_{2}\right)}\left(\hat{\mathbf{n}}_{1},\hat{\mathbf{n}}_{2}\right),
\label{eq:rotaion_biposh}
\end{align}
where $D_{m^{\prime}m}^{\ell}\left(R\right)$ is the Wigner D-matrix corresponding to the rotation. In our analysis, we evaluate the dot products in the helicity basis, in which case the rotation matrices become $R^{\prime-1}=A(\hat{\mathbf{n}})^{*} R^{-1} A(\hat{\mathbf{n}})^{T}$, where the matrix $A$ transforms between the Cartesian and helicity bases and has elements $A_{ij}=\mathbf{e}_j\cdot\mathbf{e}^{\prime}_j$, and the superscript $T$ represents a transpose, while an asterisk as a superscript denotes complex conjugation.

\subsection{Sensitivity kernels}

The fundamental measurement in time-distance helioseismology is line-of-sight projected Doppler velocity measurements of oscillations at the solar surface. We denote the time-evolution of the velocity of oscillation at the point $\mathbf{x}$ by $\mathbf{v}(\mathbf{x}, t)$. We assume that the $\mathbf{e}_x$ direction coincides with the direction of the line of sight, so the component of velocity that is measured is $\mathbf{e}_x\cdot \mathbf{v}(\mathbf{x}, t)$  The frequency-domain components of the time evolution of the velocity may be denoted by $\mathbf{v}_\omega(\mathbf{x})$. Propagation of seismic waves are best studied in terms of the cross-covariance of wave velocity at two points $\mathbf{x}_1$ and $\mathbf{x}_2$, which is given by

\begin{align}
\mathbf{C}_{\omega}\left(\mathbf{x}_{1},\mathbf{x}_{2}\right)&=\left\langle \mathbf{v}_{\omega}^{*}\left(\mathbf{x}_{1}\right)\mathbf{v}_{\omega}\left(\mathbf{x}_{2}\right)\right\rangle,
\end{align}
where the angular brackets indicate an ensemble average. The measured cross-covariance accounting for projection may be expressed as 

\begin{align}
    C_{\omega}\left(\mathbf{x}_{1},\mathbf{x}_{2}\right)&=\mathbf{e}_x\mathbf{e}_x:\mathbf{C}_{\omega}\left(\mathbf{x}_{1},\mathbf{x}_{2}\right).
\end{align}

While the cross-covariances of the radial components of the wave velocity depends solely on the angle between the two observation points $\mathbf{x}_1$ and $\mathbf{x}_2$, the line-of-sight projection breaks this spherical symmetry. We chose a second pair of points $\mathbf{x}^{\prime}_1$ and $\mathbf{x}^{\prime}_2$ that are related to the pair $(\mathbf{x}_1,\mathbf{x}_2)$ through a rotation $R$. The cross-covariances measured at these pairs of points are related through

\begin{align}
    C_{\omega}\left(\mathbf{x}_{1}^{\prime},\mathbf{x}_{2}^{\prime}\right)&=\left(R^{-1}\mathbf{e}_x\right)\left(R^{-1}\mathbf{e}_x\right):\mathbf{C}_{\omega}\left(\mathbf{x}_{1},\mathbf{x}_{2}\right).
\end{align}
The knowledge of the covariance tensor $\mathbf{C}_{\omega}\left(\mathbf{x}_{1},\mathbf{x}_{2}\right)$ therefore suffices to evaluate the line-of-sight-projected cross-covariance $C_{\omega}\left(\mathbf{x}_{1}^{\prime},\mathbf{x}_{2}^{\prime}\right)$ between any two pairs of points related by a rotation.

Following \citet{2020ApJ...905...59B}, we express the flow velocity within the Sun as

\begin{align}
    \mathbf{u}\left(\mathbf{x}\right)&=\sum_{\ell m\gamma}u_{\ell m}^{\gamma}\left(r\right)\mathbf{P}_{\ell m}^{\gamma}\left(\hat{\mathbf{n}}\right).
    \label{eq:u_3D}
\end{align}
Seismic waves interact with this velocity field, and their local propagation speeds are altered. This shows up in the measured cross-covariances $\mathbf{C}_{\omega}\left(\mathbf{x}_{1},\mathbf{x}_{2}\right)$. It is common in time-distance helioseismology to apply a filter and follow the propagation of a subsection of waves; for example, those that arrive in a short time interval. This is achieved by choosing an appropriate window that filters out wave arrivals at a particular observation point. We may obtain the difference in the time taken by seismic waves to travel between the observation points $\mathbf{x}_{1}$ and $\mathbf{x}_{2}$ by comparing the measured wave covariance with that in a model that does not account for the flow. \citet{2002ApJ...571..966G} demonstrated a way to relate the shift in travel times linearly to the flow velocity as
\begin{align}
    \delta\tau\left(\mathbf{x}_{1},\mathbf{x}_{2}\right)=\int d\mathbf{x}\,\mathbf{K}\left(\mathbf{x},\mathbf{x}_{1},\mathbf{x}_{2}\right)\cdot\mathbf{u}\left(\mathbf{x}\right).
    \label{eq:dt_3D}
\end{align}
We may use Equation \eqref{eq:u_3D} and expand the sensitivity kernel $\mathbf{K}\left(\mathbf{x},\mathbf{x}_{1},\mathbf{x}_{2}\right)$ analogously on a basis of the PB VSH to rewrite Equation \eqref{eq:dt_3D} as
\begin{align}
\delta\tau\left(\mathbf{x}_{1},\mathbf{x}_{2}\right)=\sum_{\ell m \gamma}\int_0^{R_\odot} r^{2}dr\,K_{\ell m\gamma}\left(r,\mathbf{x}_{1},\mathbf{x}_{2}\right)u_{\ell m}^{\gamma}\left(r\right).    
\end{align}
The radial profiles of the spherical harmonic coefficients of the sensitivity kernel, following \cite{2020ApJ...905...59B}, may be expressed as

\begin{align}
    K_{\gamma,\ell m}\left(r,\mathbf{x}_{1},\mathbf{x}_{2}\right)=&\int_{0}^{\infty}\frac{d\omega}{2\pi}\sum_{j_{1}j_{2}}\sum_{\alpha_{1}\alpha_{2}}K_{\ell j_{1}j_{2}\omega;\alpha_{1}\alpha_{2}}^{\gamma}\left(r,\mathbf{x}_{1},\mathbf{x}_{2}\right)\times\\\nonumber
    &\quad\mathbf{e}_x\mathbf{e}_x:\mathbf{B}_{\ell m}^{\left(j_{1}\alpha_{1}\right)\left(j_{2}\alpha_{2}\right)}\left(\hat{\mathbf{n}}_{1},\hat{\mathbf{n}}_{2}\right),
    \label{eq:kernel_radial}
\end{align}
where $j_1$ and $j_2$ are wave numbers associated with the waves, and $\alpha_1$ and $\alpha_2$ indicate vector components of the wave velocity in the helicity basis. In practice, we may limit the frequency integral to the range corresponding to waves that are bound within the Sun, although we may include higher frequencies corresponding to evanescent waves by choosing appropriate boundary conditions \citep{Gizon2017A&A...600A..35G}. The function $K_{\ell j_{1}j_{2}\omega;\alpha_{1}\alpha_{2}}^{\gamma}\left(r,\mathbf{x}_{1},\mathbf{x}_{2}\right)$ may be expressed as a product of a radial function of Green's function, which we refer to as $C_{\ell j_{1}j_{2}\omega;\alpha_{1}\alpha_{2}}^{\gamma}\left(r,r_{1},r_{2}\right)$, and a window function $h_{\omega}\left(\mathbf{r}_{1},\mathbf{r}_{2}\right)$ that selects an appropriate wave arrival, as

\begin{align}
    K_{\ell j_{1}j_{2}\omega;\alpha_{1}\alpha_{2}}^{\gamma}\left(r,\mathbf{x}_{1},\mathbf{x}_{2}\right)=2\Re\left[h_{\omega}^{*}\left(\mathbf{x}_{1},\mathbf{x}_{2}\right)C_{\ell j_{1}j_{2}\omega;\alpha_{1}\alpha_{2}}^{\gamma}\left(r,r_{1},r_{2}\right)\right].
\end{align}
The detailed expressions for these radial functions may be found in \citet{2020ApJ...905...59B}. In this work, we focused on the angular dependence of the kernel.

The window function \citep{2002ApJ...571..966G} is a nonlinear function of the modeled cross-covariance, which, in turn, implies that the kernel components $K_{\gamma,\ell m}\left(r,\mathbf{x}_{1},\mathbf{x}_{2}\right)$ and $K_{\gamma,\ell m}\left(r,\mathbf{x}^{\prime}_{1},\mathbf{x}^{\prime}_{2}\right)$ are related to each other nonlinearly through the rotation operator $R$. We may, however, split the evaluation of the kernel into two parts by noting that the components of the bipolar harmonics $\mathbf{B}_{\ell m}^{\left(j_{1}\alpha_{1}\right)\left(j_{2}\alpha_{2}\right)}\left(\hat{\mathbf{n}}_{1},\hat{\mathbf{n}}_{2}\right)$ and $\mathbf{B}_{\ell m}^{\left(j_{1}\alpha_{1}\right)\left(j_{2}\alpha_{2}\right)}\left(\hat{\mathbf{n}}^{\prime}_{1},\hat{\mathbf{n}}^{\prime}_{2}\right)$ transform linearly between each other following Equation \eqref{eq:rotaion_biposh}. The window function may be evaluated as a function of frequency while the remaining sums over $j_1$, $j_2$, $\alpha_1$ and $\alpha_2$ may be evaluated separately, and the two may be multiplied at the final stage and integrated over frequency to obtain the kernel components.

\section{Method}

We evaluated the vector harmonics by noting that these may be generated by coupling unit vectors and scalar spherical harmonics through the mechanism of angular momentum addition. Specifically, we constructed the following linear combination of the Cartesian basis vectors:

\begin{align}
    \bm{\chi}_{-1}      &=-\frac{1}{\sqrt{2}}\left(\mathbf{e}_{x}+i\mathbf{e}_{y}\right),\\
\bm{\chi}_{0}   &=\mathbf{e}_{z},\\
\bm{\chi}_{+1}  &=\frac{1}{\sqrt{2}}\left(\mathbf{e}_{x}-i\mathbf{e}_{y}\right).
\end{align}
We followed \citet{1988qtam.book.....V} and refer to this basis as the covariant spherical basis. We coupled this basis with scalar spherical harmonics to construct the vector spherical harmonics:
\begin{align}
\mathbf{Y}_{jm}^{\ell}\left(\hat{\mathbf{n}}\right) & =\sum_{\gamma=-1}^{1}\sum_{n=-\ell}^{\ell}C_{\ell n1\mu}^{jm}Y_{\ell n}\left(\hat{\mathbf{n}}\right)\bm{\chi}_{\mu}.
\label{eq:vsh}
\end{align}
The Clebsch-Gordan relations involved in this sum may be evaluated explicitly, and we list these in Appendix \ref{sect:vsh_components}. For a detailed discussion on these harmonics, we invite the reader to consult \citet{1976RSPTA.281..195J}.

We may show \citep{1988qtam.book.....V} that these harmonics are related to the Hansen ones through

\begin{align}
\mathbf{H}_{jm}^{\left(1\right)}\left(\hat{\mathbf{n}}\right) & =\sqrt{\frac{j+1}{2j+1}}\mathbf{Y}_{jm}^{j-1}\left(\hat{\mathbf{n}}\right)+\sqrt{\frac{j}{2j+1}}\mathbf{Y}_{jm}^{j+1}\left(\hat{\mathbf{n}}\right),\\
\mathbf{H}_{jm}^{\left(0\right)}\left(\hat{\mathbf{n}}\right) & =\mathbf{Y}_{jm}^{j}\left(\hat{\mathbf{n}}\right),\\
\mathbf{H}_{jm}^{\left(-1\right)}\left(\hat{\mathbf{n}}\right) & =\sqrt{\frac{j}{2j+1}}\mathbf{Y}_{jm}^{j-1}\left(\hat{\mathbf{n}}\right)-\sqrt{\frac{j+1}{2j+1}}\mathbf{Y}_{jm}^{j+1}\left(\hat{\mathbf{n}}\right).
\end{align}
This provides us with a route to get from scalar spherical harmonics to the vector ones that we are interested in. For numerical efficiency, we used the conjugation relation 
\begin{align}
\mathbf{Y}_{jm}^{\ell*}\left(\hat{\mathbf{n}}\right) & =\left(-1\right)^{j+\ell+m+1}\mathbf{Y}_{j-m}^{\ell}\left(\hat{\mathbf{n}}\right)
\end{align}
to evaluate the harmonics for negative $m$.

We developed a Julia implementation of this algorithm to evaluate vector spherical harmonics, which is available freely under the MIT license as the package VectorSphericalHarmonics.jl\footnote{\url{https://github.com/jishnub/VectorSphericalHarmonics.jl}}. In this, we used the spherical harmonics code by \citet{Limpanuparb2014arXiv1410.1748L}, which can generate accurate values of the harmonics till angular degrees of around $1000$, with an absolute and relative accuracy of $10^{-10}$. Since the components of the vector harmonics are linear combinations of normalized spherical harmonics, the accuracy of these functions are similar. This approach of computing vector spherical harmonics differs from the previous approach by \citet{2020ApJ...905...59B}, where the authors had chosen to evaluate them through the computation of Wigner D-matrices, where the accuracy was limited to degrees of around $100$. The present approach offers accuracy to a much higher degree, as well as being computationally more efficient.

The monopolar scalar or vector harmonics may be coupled to evaluate bipolar harmonics. We recognize that the expressions for the sensitivity kernels may be cast in a way that involves the harmonics 
$B_{\ell m,xx}^{\left(j_{1}\alpha_{1}\right)\left(j_{2}\alpha_{2}\right)}\left(\hat{\mathbf{n}}_{1},\hat{\mathbf{n}}_{2}\right)$ as well as $B_{\ell m,xx}^{\left(j_{1}\alpha_{1}\right)\left(j_{2}\alpha_{2}\right)}\left(\hat{\mathbf{n}}_{2},\hat{\mathbf{n}}_{1}\right)$. We may use the symmetries of the Clebsch-Gordan coefficients to obtain

\begin{align}
B_{\ell m,xx}^{\left(j_{1}\alpha_{1}\right)\left(j_{2}\alpha_{2}\right)}\left(\hat{\mathbf{n}}_{1},\hat{\mathbf{n}}_{2}\right) & =\left(-1\right)^{j_{1}+j_{2}-\ell}B_{\ell m,xx}^{\left(j_{2}\alpha_{2}\right)\left(j_{1}\alpha_{1}\right)}\left(\hat{\mathbf{n}}_{2},\hat{\mathbf{n}}_{1}\right).
\end{align}
We used this relation to avoid the need to evaluate certain bipolar harmonics, and we used pre-computed values instead.

We developed a Julia code to evaluate bipolar spherical harmonics that is freely available under the MIT license as the package BipolarSphericalHarmonics.jl\footnote{\url{https://github.com/jishnub/BipolarSphericalHarmonics.jl}}. We evaluated the Clebsch-Gordan coefficients using the freely available FORTRAN package SHTOOLS \citep{doi:10.1029/2018GC007529} for low angular degrees, and the Julia implementation of the public code wigxjpf \citep{Johansson2016} for high angular degrees. The former is computationally more efficient, using a double precision implementation of the algorithm presented by \citet{PhysRevE.57.7274} and accurate up to angular degrees of around $160$, whereas the latter is accurate at higher angular degrees as it carries out computation using arbitrary-precision arithmetic. The Julia implementation of the wigxjpf algorithm is available as the package WignerSymbols.jl. The ability to evaluate the Clebsch-Gordan coefficients for arbitrarily high degrees is an improvement over previous work by \citet{2020ApJ...905...59B}, as the kernels may now be computed using a wider range of wave modes.

We followed \citet{PhysRevE.92.043307} to evaluate Wigner D-matrices that feature in Equation \eqref{eq:rotaion_biposh}. A Julia implementation of this code is available freely under the MIT license as the package WignerD.jl\footnote{\url{https://github.com/jishnub/WignerD.jl}}. This uses an exact diagonalization of the angular momentum operator $J_y$ to evaluate the matrix elements, an approach that is tested to be accurate in absolute values up to degrees of around $100$, although relative errors may be dominated by machine precision for values close to zero. The latter does not pose a significant challenge in our analysis, given that the harmonics are normalized, and the rotated harmonics depend overwhelmingly on the D-matrix elements with absolute values on the order of $1$. Further tests might be necessary to use the rotation relation for bipolar harmonics at higher angular degrees.

\section{Results}
\begin{figure*}
    \includegraphics[width=17cm]{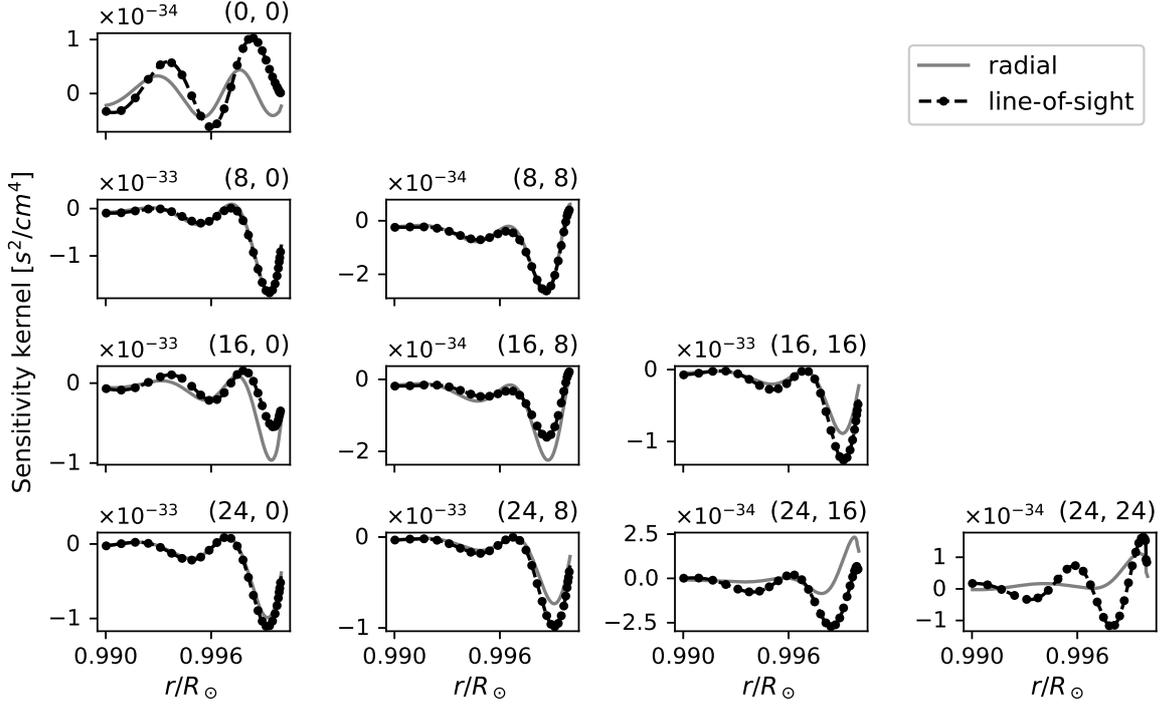}
    \caption{Radial profiles of spherical-harmonic components of the sensitivity kernel for the radial component ($\gamma=0$) of flow velocity. The gray line is computed using radial components of wave velocity, whereas the black doted line is computed using the line-of-sight projected components. The title indicates the mode indices $(\ell,m)$.}
    \label{fig:kernel_comparison}
\end{figure*}
\subsection{Sensitivity kernels}

We compared the sensitivity kernel for the radial component of the flow velocity computed using line-of-sight wave velocities and compared it with that computed using a radial component of wave velocities, choosing the two observation points to lie on the equator at a height of $150$ km above the photosphere, at azimuths of $30$ and $90$ degrees. We plot the result in Fig. \ref{fig:kernel_comparison}. The kernels were computed using Green's functions with angular degrees from $1$ to $100$, and frequencies from $2$ to $4.5$ mHz. The limitation on angular degrees arises from numerical artifacts in evaluating Green's functions using the finite-difference scheme that was presented by \citet{2020ApJ...895..117B} that arise at higher degrees; however, this is not a fundamental limitation of this approach as Green's functions computed using alternate numerical schemes may be used in its place to increase the accuracy (e.g., the finite-element scheme described by \citet{Gizon2017A&A...600A..35G}).

We see that the components of the kernels may be significantly different if line-of-sight projections are not considered. While a simple relation between the differences in the kernels and the angular distance from the disk center is difficult to obtain, our tests indicate that the difference is significant for observation points closer to the solar limb, where the line-of-sight projection effects are increasingly important to account for.

\subsection{Rotation of coordinates}

Given two sets of points $(\hat{\mathbf{n}}_1,\hat{\mathbf{n}}_2)$ and $(\hat{\mathbf{n}}^\prime_1,\hat{\mathbf{n}}^\prime_2)$ that are related through a rotation, the bipolar spherical harmonics evaluated at these points are related through Equation \eqref{eq:rotaion_biposh}. We made use of this in reverse by treating it as an active rotation of the points instead of as a passive rotation of coordinate frames. The Euler angles may be computed by expressing the rotation in terms of the points. We note that for the rotation $R,$  maps $(\hat{\mathbf{n}}_1,\hat{\mathbf{n}}_2)$ to $(\hat{\mathbf{n}}^\prime_1,\hat{\mathbf{n}}^\prime_2)$ may be expressed as the product of two rotations, where the first --- referred to as $R_1$ --- maps $\hat{\mathbf{n}}_1$ to $\hat{\mathbf{n}}^\prime_1$, and the second maps $R_1 \hat{\mathbf{n}}_2$ to $\hat{\mathbf{n}}^{\prime}_2$. The operation of the first rotation may be represented as
\begin{align}
    R_1\hat{\mathbf{n}}_{1}&=\hat{\mathbf{n}}_{1}^{\prime},\\
    R_1\hat{\mathbf{n}}_{2}&=\hat{\mathbf{n}}_{2}^{\prime\prime},
\end{align}
whereas the second one --- referred to as $R_2$ --- performs the transformation
\begin{align}
    R_2\hat{\mathbf{n}}_{2}^{\prime\prime}&=\hat{\mathbf{n}}_{2}^{\prime}.
\end{align}
The point $\hat{\mathbf{n}}^{\prime\prime}_2$ is an intermediate position that depends on the choice of $R_1$. As an aside, we note that the choice of $R_1$ is not unique. To illustrate this, we denote a rotation in the axis-angle formulation for an axis $\hat{\mathbf{n}}$ and angle $\gamma$ as $R(\hat{\mathbf{n}},\gamma)$. We see that the rotation $R^{\prime}_1 = R\left(\hat{\mathbf{n}}^\prime_1,\psi\right) R_1 R\left(\hat{\mathbf{n}}_1,\phi\right)$ for arbitrary $\psi,$ and $\phi$ also map $\hat{\mathbf{n}}_1$ to $\hat{\mathbf{n}}^\prime_1$. One choice for the rotation $R_1$ is $R_{z}\left(\phi_{1}^{\prime}\right)R_{y}\left(\theta_{1}^{\prime}-\theta_{1}\right)R_{z}\left(-\phi_{1}\right)$.

The rotation $R_2$ may be denoted in the axis-angle formulation as $R_2 = R\left(\hat{\mathbf{n}}^\prime_1,\omega\right)$, where $\omega$ satisfies

\begin{align}
\left|\hat{\mathbf{n}}_{1}^{\prime}\times\hat{\mathbf{n}}_{2}^{\prime\prime}\right|^{2}\sin\omega       &=\hat{\mathbf{n}}_{2}^{\prime}\cdot\left(\hat{\mathbf{n}}_{1}^{\prime}\times\hat{\mathbf{n}}_{2}^{\prime\prime}\right),\\
\left|\mathbf{n}_{2}^{\prime\prime}-\left(\hat{\mathbf{n}}_{1}^{\prime}\cdot\hat{\mathbf{n}}_{2}^{\prime\prime}\right)\hat{\mathbf{n}}_{1}^{\prime}\right|^{2}\cos\omega        &=\left[\mathbf{n}_{2}^{\prime}-\left(\hat{\mathbf{n}}_{1}^{\prime}\cdot\hat{\mathbf{n}}_{2}^{\prime}\right)\hat{\mathbf{n}}_{1}^{\prime}\right]\cdot\hat{\mathbf{n}}_{2}^{\prime\prime}.
\end{align}
Collectively, the rotation $R_2 R_1$ maps the set of points $(\hat{\mathbf{n}}_1,\hat{\mathbf{n}}_2)$ to $(\hat{\mathbf{n}}^\prime_1,\hat{\mathbf{n}}^\prime_2)$. Given this rotation, we may use Equation \eqref{eq:rotaion_biposh} to evaluate the bipolar harmonics $\mathbf{B}_{\ell m}^{\left(j_{1}\alpha_{1}\right)\left(j_{2}\alpha_{2}\right)}\left(\hat{\mathbf{n}}^{\prime}_{1},\hat{\mathbf{n}}^{\prime}_{2}\right)$ by transforming $\mathbf{B}_{\ell m}^{\left(j_{1}\alpha_{1}\right)\left(j_{2}\alpha_{2}\right)}\left(\hat{\mathbf{n}}_{1},\hat{\mathbf{n}}_{2}\right)$ instead of an explicit evaluation.

\begin{figure}
    \includegraphics[scale=0.75]{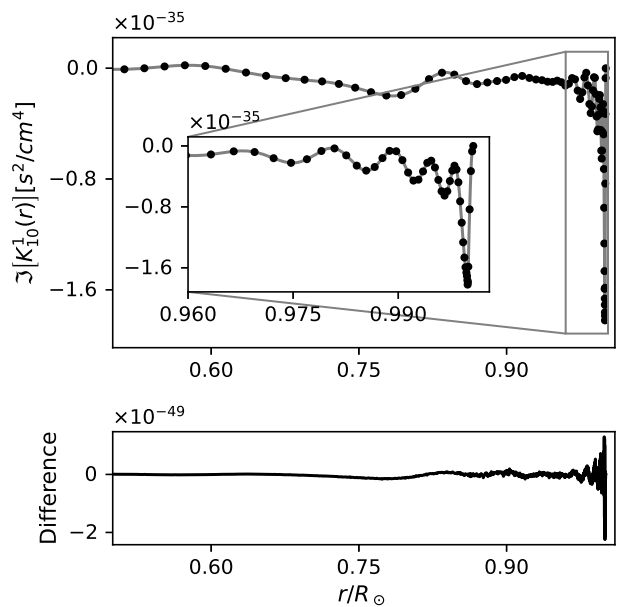}
    \caption{Radial profiles of the imaginary part of the kernel for $(\ell=1,m=0)$ and vector index $\gamma=1$, computed directly for $\hat{\mathbf{n}}_1=(\pi/2,0)$ and $\hat{\mathbf{n}}_2=(\pi/2,\pi/4)$ using the equation (solid line) and by rotating the kernels for $\hat{\mathbf{n}}_1=(0,0)$ and $\hat{\mathbf{n}}_2=(\pi/4,0)$ (dots). The kernels are computed for line-of-sight projected measurements. The bottom panel represents the absolute difference between the two results.}
    \label{fig:rotpts}
\end{figure}

We demonstrate the transformation of sensitivity kernels in Fig. \ref{fig:rotpts} by choosing the pairs of points $\hat{\mathbf{n}}_1=(0, 0)$, $\hat{\mathbf{n}}_2=(\pi/4, 0)$, $\hat{\mathbf{n}}^\prime_1=(\pi/2, 0)$ and $\hat{\mathbf{n}}^\prime_2=(\pi/2, \pi/4)$. The solid line represents the kernels computed using Equation \eqref{eq:kernel_radial} for the points $\hat{\mathbf{n}}^\prime_1$ and $\hat{\mathbf{n}}^\prime_2$, while the dots represent that computed by rotating the kernel components for $\hat{\mathbf{n}}_1$ and $\hat{\mathbf{n}}_2$ using Equation \eqref{eq:rotaion_biposh}. The bottom panel of Fig. \ref{fig:rotpts} depicts the absolute difference between the kernels computed using the two approaches, and the magnitude of the difference indicates that the error is almost purely numerical. The close match shows that the components of sensitivity kernels for line-of-sight projected wave velocities for points that are related through a rotation may be transformed among each other, and kernels for several pairs of points may be computed simultaneously. In the presence of center-to-limb factors that break spherical symmetry (e.g., differences in line-formation heights), such a transformation may only be carried out among pairs of points for which symmetry-breaking factors remain unchanged on rotation.

\subsection{Evaluation time}

\begin{figure}
    \centering
    % \resizebox{\hsize}{!}{
    %     \includegraphics{runtimes.eps}
    % }
    
    \includegraphics[scale=0.75]{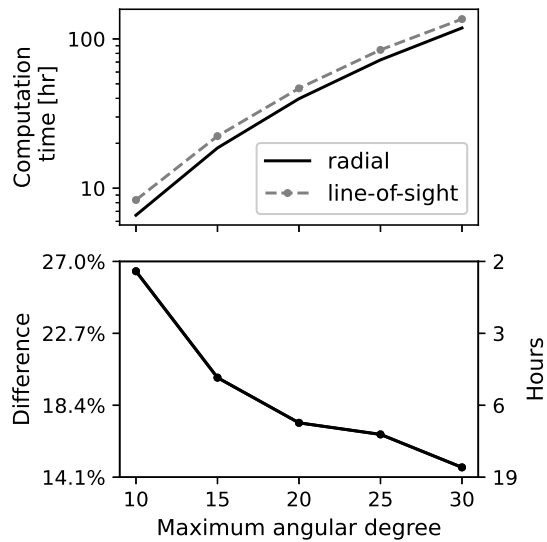}
    
    \caption{Computational time required to evaluate kernels $K_{\ell m}$ for all modes for $0\leq \ell \leq \ell_\text{max}$ and $0 \leq m \leq \ell_\text{mmax}$, where $\ell_\text{max}$ is the angular degree cut-off. In the top panel, the solid line indicates the time required using the radial component of the displacement, while the dashed line indicates the time required using the line-of-sight projected component of the wave velocity.}
    \label{fig:runtimes}
\end{figure}

The numerical implementation of the algorithm described in \citet{2020ApJ...905...59B} has three parts --- the first where the angular functions are computed, the second where the radial functions are evaluated, and the third where the products of the two are summed over. In this work, we described an approach to perform the first part more efficiently than before. We also improved upon the second and third parts by harnessing Julia's loop vectorization capabilities. We demonstrate one comparison of the evaluation times in Fig. \ref{fig:runtimes}, where we plot the time required to evaluate the kernels using the line-of-sight projections of wave velocity and that using the radial component of wave velocity with the maximum angular degree up to which which the kernels are evaluated. We used Green's functions corresponding to wave numbers in the $1$ to $100$ range and frequency in the $2$ mHz to $4.5$ mHz range uniformly divided over $4000$ bins.  The relative difference between the evaluation times using line-of-sight projected velocities and that using radial velocities decreases with an increase in the cutoff in angular degree, reaching $15\%$ at $\ell_\text{max}=30$. The absolute difference using $\ell_\text{max}=30$ is $17.5$ hours of computation time. The timings were measured on $224$ Intel Xeon CPU E5-2680 v4 cores running at $2.40$GHz.

\section{Discussion and conclusions}

In this work, we improved upon the previous work by \citet{2020ApJ...905...59B} to evaluate helioseismic sensitivity kernels for travel-time shifts computed using line-of-sight projected seismic wave velocities. The ability to transform between kernels for pairs of points that are related by a rotation may be helpful if one is interested in inverting for averaged travel times to improve the signal-to-noise ratio. The Julia implementation of the approach presented in this work is available freely under the MIT license \footnote{\url{https://github.com/jishnub/HelioseismicKernels.jl}}. The modular structure of the code ensures that parts of the entire project, such as vector spherical harmonics, bipolar spherical harmonics, or Wigner D-matrices, may be reused by other projects as necessary. The code developed is embarrassingly parallel in wave modes and temporal frequency and uses the message passing interface (MPI) for communication between processes, so it may be directly run on computing clusters to evaluate sensitivity kernels.

In this work, we did not focus on the impact of the differences in line-formation heights on the sensitivity kernels, although it is expected to play a part in observed travel-time differences as well \citep{2020ASSP...57..123Z}. While this may also be incorporated into our approach, it requires a more careful modeling of the observation heights and the atmospheric boundary condition experienced by seismic waves. Further refinements in this direction might be possible. We also assumed that the wave excitation occurs at one specific radius, while in reality the excitation is perhaps spread over a radial range. A different model of source covariance following \citet{Gizon2017A&A...600A..35G} might be used to obtain a better match to observations, although this would significantly increase the computational burden. Furthermore, the temporal spectrum of the source covariance was chosen arbitrarily in this work to be a Gaussian centered at $3.2$ mHz, although a better profile may be obtained from the spectral envelope of the observed wave covariances.

\begin{acknowledgements}
This work was supported by NYUAD Institute Grant
G1502 "NYUAD Center for Space Science".
This research was carried out on the High-Performance Computing resources at New York University Abu Dhabi. 

\end{acknowledgements}
\bibliographystyle{aa}
\bibliography{references}
\newpage
\begin{appendix}
\section{Vector spherical harmonics\label{sect:vsh_components}}
We numerically evaluated the vector spherical harmonics following Equation \eqref{eq:vsh}. We list the components of the VSH in the spherical basis, as detailed by \citet{1988qtam.book.....V}:

\begin{align}
    \left[\mathbf{Y}_{jm}^{j+1}\left(\hat{\mathbf{n}}\right)\right]^{+1}&=\left[\frac{\left(j-m+1\right)\left(j-m+2\right)}{2\left(j+1\right)\left(2j+3\right)}\right]^{\frac{1}{2}}Y_{j+1m-1}\left(\hat{\mathbf{n}}\right),\\
    \left[\mathbf{Y}_{jm}^{j+1}\left(\hat{\mathbf{n}}\right)\right]^{0}&=-\left[\frac{\left(j-m+1\right)\left(j+m+1\right)}{\left(j+1\right)\left(2j+3\right)}\right]^{\frac{1}{2}}Y_{j+1m}\left(\hat{\mathbf{n}}\right),\\
    \left[\mathbf{Y}_{jm}^{j+1}\left(\hat{\mathbf{n}}\right)\right]^{-1}&=\left[\frac{\left(j+m+1\right)\left(j+m+2\right)}{2\left(j+1\right)\left(2j+3\right)}\right]^{\frac{1}{2}}Y_{j+1m+1}\left(\hat{\mathbf{n}}\right),\\
    \left[\mathbf{Y}_{jm}^{j}\left(\hat{\mathbf{n}}\right)\right]^{+1}&=-\left[\frac{\left(j+m\right)\left(j-m+1\right)}{2j\left(j+1\right)}\right]^{\frac{1}{2}}Y_{jm-1}\left(\hat{\mathbf{n}}\right),\\
    \left[\mathbf{Y}_{jm}^{j}\left(\hat{\mathbf{n}}\right)\right]^{0}&=\frac{m}{\sqrt{j\left(j+1\right)}}Y_{jm}\left(\hat{\mathbf{n}}\right),\\
    \left[\mathbf{Y}_{jm}^{j}\left(\hat{\mathbf{n}}\right)\right]^{-1}&=\left[\frac{\left(j-m\right)\left(j+m+1\right)}{2j\left(j+1\right)}\right]^{\frac{1}{2}}Y_{jm+1}\left(\hat{\mathbf{n}}\right),\\
    \left[\mathbf{Y}_{jm}^{j-1}\left(\hat{\mathbf{n}}\right)\right]^{+1}&=\left[\frac{\left(j+m\right)\left(j+m-1\right)}{2j\left(2j-1\right)}\right]^{\frac{1}{2}}Y_{j-1m-1}\left(\hat{\mathbf{n}}\right),\\
    \left[\mathbf{Y}_{jm}^{j-1}\left(\hat{\mathbf{n}}\right)\right]^{0}&=\left[\frac{\left(j-m\right)\left(j+m\right)}{j\left(2j-1\right)}\right]^{\frac{1}{2}}Y_{j-1m}\left(\hat{\mathbf{n}}\right),\\
    \left[\mathbf{Y}_{jm}^{j-1}\left(\hat{\mathbf{n}}\right)\right]^{-1}&=\left[\frac{\left(j-m\right)\left(j-m-1\right)}{2j\left(2j-1\right)}\right]^{\frac{1}{2}}Y_{j-1m+1}\left(\hat{\mathbf{n}}\right).
\end{align}
The spherical harmonics $Y_{\ell m}(\hat{\mathbf{n}})$ are zero for indices $(\ell,m)$ that do not satisfy $\left|m\right|\leq\ell$.

\end{appendix}

\end{document}